\documentclass{osa-article}

\journal{oe}

\articletype{Research Article}
\usepackage{bm}
\usepackage{graphicx}
\usepackage{subfigure}
\usepackage[noend]{algpseudocode}
\usepackage[ruled]{algorithm2e}
\begin{document}

\title{Single-mode fiber coupling with a M-SPGD algorithm for long-range quantum communications} 

\author{Kui-Xing Yang,\authormark{1,2,3,6} Maimaiti Abulizi,\authormark{1,2,4,6}Yu-huai Li,\authormark{1,2,3}Bo-Yang Zhang,\authormark{1,2,3}Shuang-Lin Li,\authormark{1,2,3} Wei-yue Liu,\authormark{1,2,5} Juan Yin,\authormark{1,2,3} Yuan Cao,\authormark{1,2,3,7} Ji-gang Ren\authormark{1,2,3,8} and Cheng-zhi Peng\authormark{1,2,3} }

\address{\authormark{1}Hefei National Laboratory for Physical Sciences at the Microscale and Department of Modern Physics, University of Science and Technology of China, Hefei 230026, China\\
\authormark{2}Shanghai Branch, CAS Center for Excellence in Quantum Information and Quantum Physics, University of Science and Technology of China, Shanghai 201315, China\\
\authormark{3}Shanghai Research Center for Quantum Sciences, Shanghai 201315, China\\
\authormark{4}School of Physics and Electronic Engineering, Xinjiang Normal University, Urumqi 830054, China\\
\authormark{5}College of Information Science and Engineering, Ningbo University, Ningbo 315211, China\\
}

\address{\authormark{6}These authors contributed equally to this work\\}
\email{\authormark{7}yuancao@ustc.edu.cn} 
\email{\authormark{8}jgren@ustc.edu.cn} 



\begin{abstract}
Satellite-based quantum communication is a promising approach for realizing global-scale quantum networks. For free-space quantum channel, single-mode fiber coupling is particularly important for improving signal-to-noise ratio of daylight quantum key distribution (QKD) and compatibility with standard fiber-based QKD. However, achieving a highly efficient and stable single-mode coupling efficiency under strong atmospheric turbulence remains experimentally challenging. Here, we develop a single-mode receiver with an adaptive optics (AO) system based on a modal version of the stochastic parallel gradient descent (M-SPGD) algorithm and test its performance over an 8 km urban terrestrial free-space channel. Under strong atmospheric turbulence, the M-SPGD AO system obtains an improvement of about 3.7 dB in the single-mode fiber coupling efficiency and a significant suppression of fluctuation, which can find its applications in free-space long-range quantum communications.
\end{abstract}

\section{Introduction}
Quantum key distribution (QKD) provides information-theoretic secure keys between two remote parties based on the laws of quantum mechanics.
Since the development of the first QKD protocol in 1984\cite{bennett1984proceedings}, quantum information processing, particularly quantum communication, has attracted a diverse field of researchers in theoretical and experimental physics\cite{gisin2007quantum, xu2020secure}. 
QKD based on satellites is treated as one of the most promising solutions to realize a global quantum communication network.
Many excellent experimental works have been performed on the ground in the past decade to demonstrate the feasibility of satellite-based QKD \cite{peng2005experimental,schmitt2007experimental,bonato2009feasibility,yin2012quantum,nauerth2013air,wang2013direct,cao2013entanglement,bourgoin2015free,vallone2015experimental,bourgoin2015experimental,bedington2016nanosatellite,pugh2017airborne,takenaka2017satellite}.
Recently, a series of ground-breaking experiments based on the Micius satellite have taken the first step toward global-scale quantum communication\cite{yin2017satellite, liao2017satellite,ren2017ground,liao2018satellite}.

An important next step is to extend the satellite-based QKD from nighttime to daytime, however, the strong background noise from the scattered sunlight is typically five orders of magnitude greater than the background noise during the night\cite{er2005background}, thus a high-efficiency single-mode fiber (SMF) coupling under strong atmospheric turbulence is necessary to provide a sufficient signal-to-noise ratio (SNR) \cite{liao2017long,avesani2019full,ko2018experimental}. 
Furthermore, additional complex quantum information processing involving quantum interference\cite{pan2012multiphoton}, such as quantum teleportation\cite{bouwmeester1997experimental}, entanglement swapping and purification\cite{pan1998experimental,pan2001entanglement}, also requires a high-efficiency SMF coupling for filtering the spatial mode. 

However, attainment of a high-efficiency SMF coupling remains technically challenging due to the effect of atmospheric turbulence.
In principle, adaptive optics (AO) provides a solution to this challenge\cite{wright2015adaptive,gong2018free}.
AO can be divided into two categories: conventional AO with aberration measurement and optimized AO without aberration measurement.    
Strong turbulence can cause intensity scintillations and phase discontinuity, which leads to error in wave-front measurement\cite{barchers2002evaluation,chen2007detection}. 
Therefore, under the condition of strong turbulence, conventional AO technology may be inefficient.
An optimized AO system does not require the wave-front reconstruction needed for conventional AO systems as it directly optimizes the performance metric using the control signal of the wave-front corrector as an optimization parameter. 
 
The optimized AO technique has been proven to be more robust and resilient than the conventional AO technique; however, it requires a higher operational bandwidth due to the iterative nature of the control algorithm\cite{weyrauch2001microscale}. 
The stochastic parallel gradient descent (SPGD) algorithm proposed by Voronstov in 1997 significantly increased the speed of wavefront distortion correction comparing with other model-free optimization methods\cite{vorontsov2000adaptive}. A preliminary attempt to employ AO based on SPGD algorithm in daylight free-space QKD experiment has been performed in 2018\cite{gong2018free}. However, due to large number of actuators in a deformable mirror, it will take hundreds of iterations to complete an effective wavefront correction, which great limits the performance of the application of AO in free-space QKD. The modal version of the SPGD algorithm (M-SPGD) presented by Martin J. Booth is a possible solution\cite{booth2006wave}, which has a higher convergence rate than the conventional SPGD. Numerous simulation and desktop demonstration experiments have illustrated the effectiveness of the M-SPGD algorithm\cite{fu2014stochastic,xie2015phase,anzuola2016performance}; however, its performance under real long-distance atmospheric conditions remains to be tested.

Here, based on M-SPGD algorithm, we develop an optimized AO system and apply it to optimize SMF coupling. We implement a field test of the performance of the M-SPGD AO system over a long-distance horizontal link, and definitely observe the enhancement of SMF coupling efficiency and the suppression of the coupling efficiency fluctuation quantitatively under different turbulence intensities. Our results further show the great potential of M-SPGD AO technology in realizing the long-range quantum communications.

\section{AO system based on the M-SPGD algorithm}

\subsection{Principle of the M-SPGD algorithm}
In SPGD algorithm, the performance metric $ J $, i.e., the coupled power of SMF in our system, is taken as the objective function of the control voltage vector $ \boldsymbol{u} $ of a deformable mirror, where the dimension of the voltage vector is equal to the number of actuators. 
The iterative update rule is given by
\begin{equation}
\boldsymbol{u}^{m+1}=\boldsymbol{u}^{m}+G \cdot \delta J^{m} \cdot \delta \boldsymbol{u}
\end{equation}
where $ m $ is the iteration step, $ G $ is the gain coefficient, and $ \delta J^{m}=J(\boldsymbol{u}^{m}+\delta\boldsymbol{u})-J(\boldsymbol{u}^{m}-\delta\boldsymbol{u}) $ is the performance metric perturbation obtained after applying a positive perturbation voltage $ \boldsymbol{u}^{m}+\delta\boldsymbol{u} $
and negative perturbation voltage $\boldsymbol{u}^{m}-\delta\boldsymbol{u}$ to the defromable mirror. 
The perturbation voltage $ \delta\boldsymbol{u} $ follows a Bernoulli distribution with zero mean. 
The amplitude of the perturbation voltage $ |\delta\boldsymbol{u}| $ and the gain coefficient $ G $ are very important parameters in SPGD algorithm. 
Among model-free optimization methods, approaches based on the stochastic gradient methods proposed by M. Voronstov have been verified as the fastest search methods\cite{vorontsov2000adaptive,linhai2011wavefront}.
\par
SPGD algorithm is essentially a rapid searching algorithm; thus, a smaller parameter space ensures a higher convergence rate. In above traditional SPGD algorithm, the number of parameters is equal to the number of actuators. In practice, the shape of the deformable mirror for wavefront aberration compensation can be decomposed by a certain mode, for example, Zernike mode, which consists of a series of orthogonal Zernike polynomials. Wave-front aberration can be expressed as $ Z(\rho, \theta)=\sum_{n=0}^{N} a_{n} \times Z_{n}(\rho, \theta) $, where $ Z_{n}(\rho, \theta) $ is the nth Zernike polynomial and $ a_{n} $ is the corresponding coefficient. So, the performance metric $ J $ can be also seen as the function of Zernike coefficient vector $ \boldsymbol{a} $, then the number of parameters is equal to the number of Zernike modes we selected, the optimization formula of M-SPGD algorithm is given by
\begin{equation}
\boldsymbol{a}^{m+1}=\boldsymbol{a}^{m}+G \cdot \delta J^{m} \cdot \delta \boldsymbol{a}
\end{equation}
where $ m $ is the iteration step, $ G $ is the gain coefficient, and $ \delta J^{m}=J(\boldsymbol{a}^{m}+\delta\boldsymbol{a})-J(\boldsymbol{a}^{m}-\delta\boldsymbol{a}) $ is the performance metric perturbation obtained after applying a positive  Zernike coefficient vector perturbation $ \boldsymbol{a}^{m}+\delta\boldsymbol{a} $ and negative  Zernike coefficient vector perturbation $\boldsymbol{a}^{m}-\delta\boldsymbol{a}$ to the deformable mirror. 
The Zernike coefficient vector perturbation $ \delta\boldsymbol{a} $ follows a Bernoulli distribution with zero mean. In our M-SPGD AO system, the number of Zernike modes used is 12 and the number of actuators of the deformable mirror is 40. In other words, the number of parameters has been reduced from 40 to 12.  
\par
\begin{algorithm}[H]
\label{m-spgd}
\caption{M-SPGD algorithm. $ N $ is number of iterations of the algorithm; $ M $ is transformation matrix for mapping Zernike coefficient vector into driving voltage vector.}
\LinesNumbered
Zernike coefficient vector $ \boldsymbol{a} $ initialization\;
counter m = 0\;
\For{m < N}{
    Zernike coefficient vector perturbation $ \delta\boldsymbol{a} $ is randomly generated \;
    driving voltages calculation according to Zernike coefficient vector perturbation:\\
    positive perturbation voltage vector $ \boldsymbol{u}^{m}_{+} $ = $ (\boldsymbol{a}^{m}+\delta\boldsymbol{a})\times\boldsymbol{M} $\;
    negative perturbation voltage vector $ \boldsymbol{u}^{m}_{-} $ = $ (\boldsymbol{a}^{m}-\delta\boldsymbol{a})\times\boldsymbol{M} $\;
    $ \boldsymbol{u}^{m}_{+} $ is applied to the deformable mirror \;
    the performance metric $ J(\boldsymbol{a}^{m}+\delta\boldsymbol{a}) $ is measured \;
    $ \boldsymbol{u}^{m}_{-} $ is applied to the deformable mirror \;
    the performance metric $ J(\boldsymbol{a}^{m}-\delta\boldsymbol{a}) $ is measured \;
    the performance metric change is calculated: $ \delta 5J^{m}=J(\boldsymbol{a}^{m}+\delta\boldsymbol{a})-J(\boldsymbol{a}^{m}-\delta\boldsymbol{a}) $\;
    the Zernike coefficient vector is updated:\\
    $ \boldsymbol{a}^{m+1}=\boldsymbol{a}^{m}+G \cdot \delta J^{m} \cdot \delta \boldsymbol{a} $ \;
    m = m + 1
}
\end{algorithm}
\par
In the Algorithm\ref{m-spgd} described above, although additional calculations for mapping the Zernike coefficients to the control voltages of the actuators are needed in the iteration, the calculation speed is usually extremely high. The phase modulated by an actuator of the deformable mirror is proportional to driving voltage, so the mapping process from Zernike coefficients to driving voltages can be realized just by multiplying a constant linear transformation matrix $ \boldsymbol{M} $. Therefore, in the M-SPGD algorithm, we can achieve a higher speed of wave-front distortion correction by choosing the number of Zernike modes. 

\subsection{Realization of an M-SPGD AO system}

In this study, we utilized a silver-coated piezoelectric deformable mirror with 40 actuators and a full stroke bandwidth of approximately 2 kHz. 
An infrared photodetector (PD) with a high gain and high bandwidth was used to detect the optical signal, and a multi-function data acquisition (DAQ) card was applied in our AO system. 
This DAQ card is used not only to read the performance metric but also for precise time-delay control in the AO closed-loop. 
To ensure that the readout of the performance metric occurs right after the random perturbation action of the deformable mirror, we designed the following system to measure the delay of the closed-loop control, as shown in Fig. 1.
\begin{figure}[htbp]
	\centering
	\includegraphics[width=10cm]{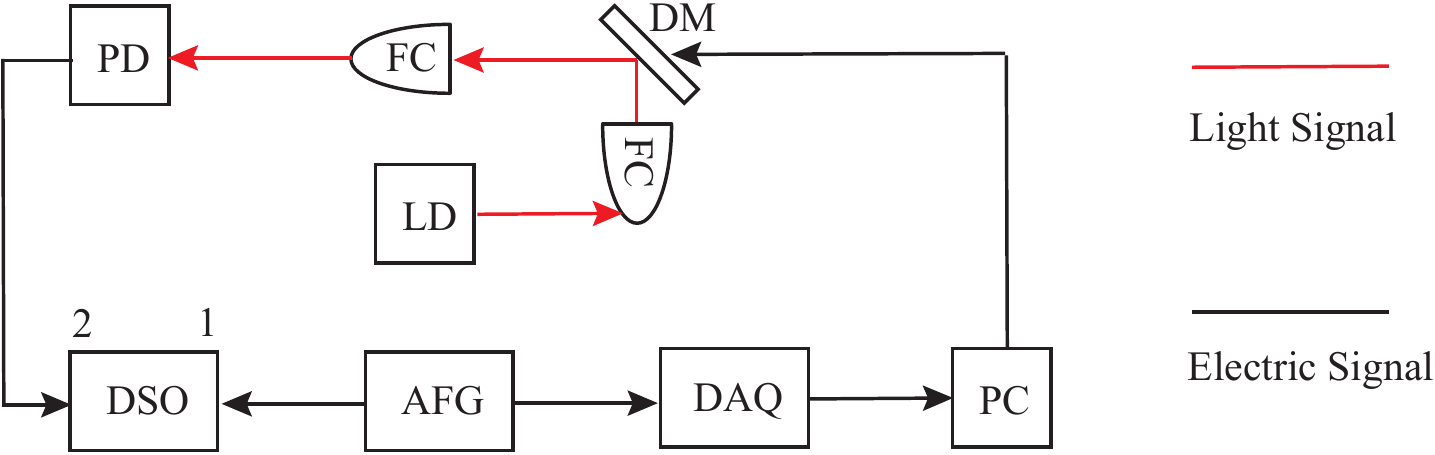}
	\caption{\textbf{Latency of the AO system measurement.} AFG, arbitrary function generator; DSO, digital storage oscilloscope; PC, personal computer; DM, deformable mirror; FC, fiber collimator; LD, laser diode;}
	\label{Fig. 1}
\end{figure}
An AFG generates two synchronous pulse signals. One signal is sent directly into the first channel of the DSO as a trigger, and the second signal is sent into the DAQ to be detected. The PC continuously sends different commands to the deformable mirror based on the signal voltage level, so different optical power coupled into the FC is obtained and the PD converts optical signals to electrical signals. The output voltage of the PD is sent into the second DSO channel. The delay of the AO closed-loop control, which is about 900 $ \mu s $, can be precisely measured. We utilize the high-speed DAQ card to accurately control the delay, and a delay of approximately 900 $ \mu s $ is applied to the readout of the performance metric after the disturbance command is issued. The iterative frequency of our AO system is approximately 500 Hz.
\par
To obtain optimal performance from the algorithm, an optimization of the parameters, particularly the disturbance step size and gain coefficient, is required. The algorithm we designed automatically attempts different combinations of disturbance step sizes and gain coefficients, and the average SMF coupled power is used as the evaluation criterion to select the optimal parameters. 
\section{SMF coupling with the M-SPGD AO system and a field test}
The SMF sending terminal Alice is located at Lane 188, Dujuan Road $ (N31^{\circ}12^{\prime}15^{\prime\prime},E121^{\circ}32^{\prime}45^{\prime\prime}) $ in Pudong New Area, and the SMF receiving terminal Bob is located at Lane 68, Xiuyan Road $(N31^{\circ}8^{\prime}2^{\prime\prime},E121^{\circ}32^{\prime}2^{\prime\prime})$ in Pudong New Area. 
The distance between the sender and receiver is approximately 8 km. 
To ensure a mutual view, the sending system is located on the 28th floor of a building, while the receiving system is located on the 11th floor of a building. 
A corresponding aerial view and schematic diagram of the terminals are shown in Fig. 2.
\begin{figure}[htbp]
	\centering
	\includegraphics[width=9cm]{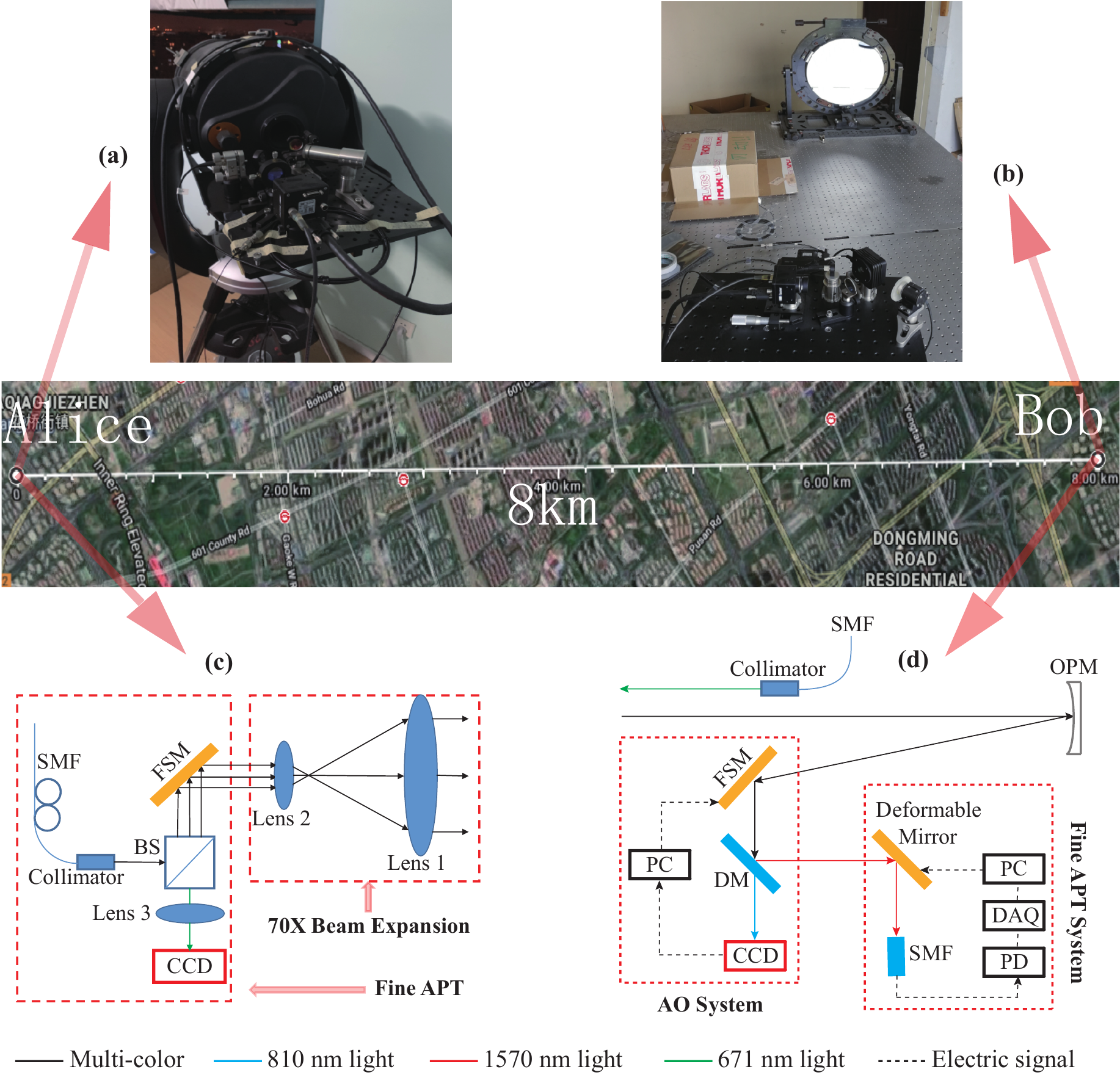}
	\caption{\textbf{Aerial view of an 8-km free-space horizontal link.} \textbf{(a)}, The physical setup of transmitter Alice. \textbf{(b)}, The physical setup of receiver Bob. \textbf{(c)}, Schematic diagram of the optics of the transmitter. The SMF transmitting telescope terminal primarily comprises two components: an acquiring, pointing, and tracking (APT) system and a beam expanding system. The APT system includes an SMF collimated laser transmitter, a high-speed industrial camera (CCD), a beam splitter (BS), and a fast steering mirror (FSM). The 70-fold beam-expanding system is composed of an achromatic lens with a focal length of 40 mm and a telescope with a focal length of 2800 mm that shares the same focal position. \textbf{(d)}, Schematic diagram of the optics of the receiver. The receiving telescope terminal comprises a fine APT system and an optimized AO system. The fine APT system includes a FSM, a dichroic mirror (DM), a high-speed CCD, and a receiving SMF, the AO system comprises a piezoelectric deformable mirror, a high-speed photodetector (PD), a high-speed data acquisition (DAQ) card, and an algorithm-realized PC. OPM, off-axial parabolic mirror.}
	\label{Fig. 2}
\end{figure}
At the transmitting terminal Alice, 810 nm and 1570 nm light beams are sent together from single-mode fiber (SMF) with operating wavelength of 1460-1620 nm and numerical aperture (NA) of 0.13 rad, then pass through a collimator with focal length 15 mm. The 810 nm laser is used as beacon light of the fine APT system of receiving terminal Bob and 1570 nm laser is used as signal light for AO performance test. The divergence angle of the signal light is carefully calibrated to be about 20 $ \mu $rad and the diameter of the signal light is about 273 mm which is smaller than the diameter of the telescope 280 mm. The fine APT system of Alice uses 671 nm laser which is transmitted from one side of Bob telescope as beacon light to ensure that the 1570 nm signal light is pointed to the receiving telescope accurately. 
At the receiving terminal Bob, 810 nm and 1570 nm lasers are collected together by an off-axial parabolic mirror with focal length 2000 mm. 
The fine APT system can achieve a tracking accuracy of 2.5 $ \mu $rad for tilt aberration correction using 810 nm beacon beam. 
In the AO system, the 1570 nm signal light is detected by a photodetector after being coupled into SMF. 
A PC is used to realize M-SPGD algorithm, which reads the output voltage of the PD by a DAQ card and outputs signals to control the defromable mirror. 
The 1570 nm and 810 nm lasers are separated by a dichromatic mirror.
\par
Based on numerous tests over an 8-km free-space link, the results show that our M-SPGD algorithm AO system can effectively improve the SMF coupling efficiency and suppress intensity fluctuations. Moreover, we also used angle-of-arrival fluctuation method to estimate the atmospheric coherence length $ r_{0} $, also called Fried parameter, in order to characterize the strength of the atmospheric turbulence\cite{martin1987image}. 
\begin{equation}
r_{0}=3.18 k^{-6 / 5} D^{-1 / 5} \delta_{\alpha}^{-6 / 5}
\end{equation}
Here, $ \delta_{\alpha}^{2} $ is the variance of angle-of-arrival fluctuation, $ D $ is the diameter of receiving telescope and $ k $ is angular wavenumber. To measure angle-of-arrival fluctuation $ \delta_{\alpha} $ at the receiving terminal Bob, APT system and AO system are both set to open-loop, CCD of the APT system is used to record image motion with a speed about 1 kHz, so angle-of-arrival fluctuation can be calculated by formula $ \delta_{\alpha}= stdev(C_{I}/f) $, where $ C_{I} $ is position of the centroid of the image on CCD, $ f $ is focal length of the OPM and $ stdev $ represents standard deviation. For evaluating AO system performance, normalized atmospheric turbulence strength $ D/r_{0} $ is more suitable to present the atmospheric conditions. 
\begin{figure}[htbp]
	\centering
	\includegraphics[width=10cm]{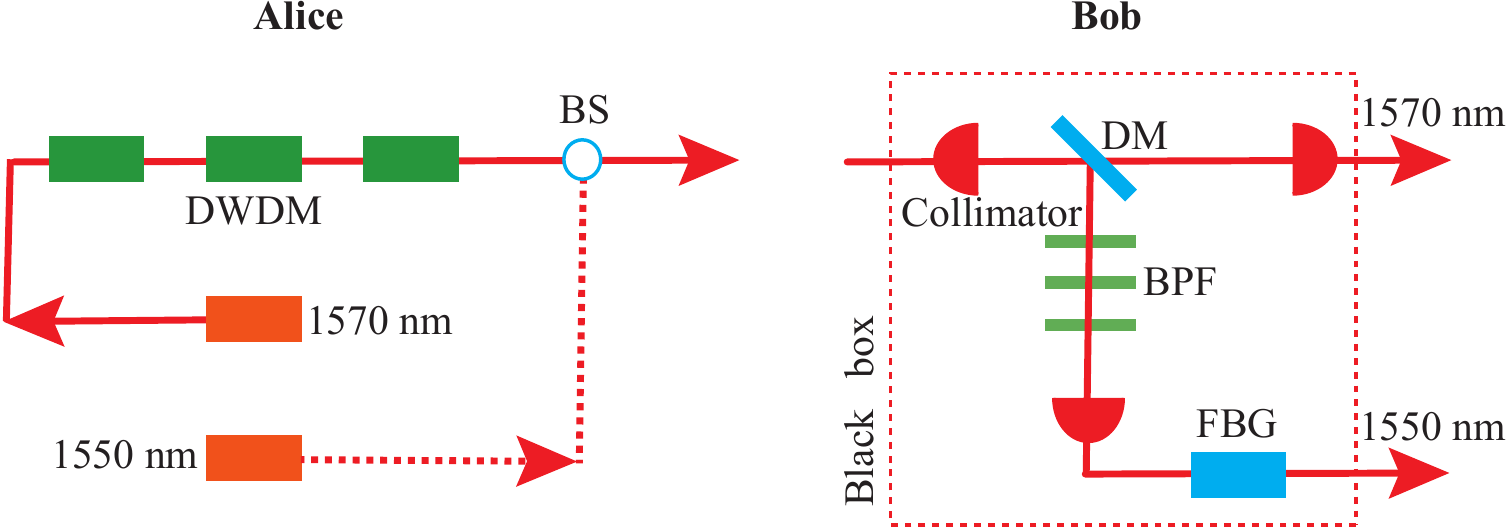}
	\caption{\textbf{Filtering setup.} DWDM, dense wavelength division multiplexing; BS, beam splitter; DM, dichroic mirror; BPF, band-pass filter; FBG, fiber bragg grating.}
	\label{Fig. 3}
\end{figure}
\par
In general, 1550 nm light is usually selected as signal light for free-space daylight quantum communication because of weaker solar background noise \cite{liao2017long,avesani2019full}, which implies that the 1570 nm light used for M-SPGD AO optimization will contribute to dark counts. We designed necessary filtering setup to reduce the noise introduced by 1570 nm light and measure the noise level after filtering. As shown in Fig. 3, Alice uses three DWDMs with a center wavelength 1570.42 nm and a bandwidth 100 GHz to reduce sideband noise of the 1570 nm light source. After collecting 1550 nm light and 1570 nm light by SMF together, Bob utilizes a DM to separate 1570 nm light and 1550 nm signal light first, 1570 nm light is used for M-SPGD AO system and 1550 nm light is used for quantum communication. Then three BPFs with a center wavelength 1550 nm  and a bandwidth 3 nm are used to reduce noise introduced by the 1570 nm light source. After being coupled into SMF, the 1550 nm signal light pass through a FBG with an ultra-narrow bandwidth 40 pm to further suppress noise. Bob's filter module is installed on a breadboard and placed in a black box to avoid additional background noise. The noise is detected by a superconducting nanowire single photon detector (SNSPD) with an efficiency of $ 80\% $ and dark counts of about 30 Hz.
\par
It should be noted that 1550 nm light was not sent from Alice really in our experiment, we just measure dark counts introduced by 1570 nm light from the 1550 nm export of the filter module to evaluate the SNR of quantum communication system. The signal light wavelength of the quantum communication system is assumed to be 1550 nm and the filter module was adjusted to guarantee maximum transmittance of the 1550 nm light. A complete QKD system based on 1550 nm light source can be directly integrated into our system in fact, for example, a BS can be used to combine 1550 nm signal light with 1570 nm light as shown in Fig. 3.

\section{Result and discussion}
The coupled 1570 nm light power which is used as objective function is detected by the high-speed photodetector, at the same time the coupling efficiency is recorded with a sample rate of 500 Hz, which is consistent with the bandwidth of the AO system.
\par
As shown in Fig. 4(a), for a normalized atmospheric coherence length of $ D/r_{0}=5.4 $ ($r_{0}=7.4$ cm @810 nm), which corresponds to moderate intensity atmospheric turbulence,  the improvement of single-mode coupling efficiency is about 3.1 dB from open-loop coupling efficiency $ 4.8\% $ to closed-loop coupling efficiency $ 9.7\% $ by using the M-SPGD AO system. For free-space quantum communication, at a certain noise level higher coupling efficiency means higher SNR, which will result in higher key rate. From the histogram of coupling efficiency statistical distribution Fig. 4(b), we can intuitively see that the coupling efficiency of closed-loop gathers to a smaller span comparing with that of open-loop. Specifically and quantitatively, relative standard deviation (RSD) is used to represent coupling efficiency fluctuation. The closed-loop single-mode coupling efficiency RSD is approximately $ 51.1\% $ while the open-loop single-mode coupling efficiency RSD is approximately $ 88.7\% $. In other words, the coupling efficiency fluctuation is obviously suppressed, which benefits quantum information processing tasks involving quantum interference, because matched intensity will result in higher interference visibility. The fluctuation suppression shows that our M-SPGD AO system can compensate dynamic wavefront distortion introduced by atmospheric turbulence.
\begin{figure}[htbp]
\centering
\includegraphics[width=9cm]{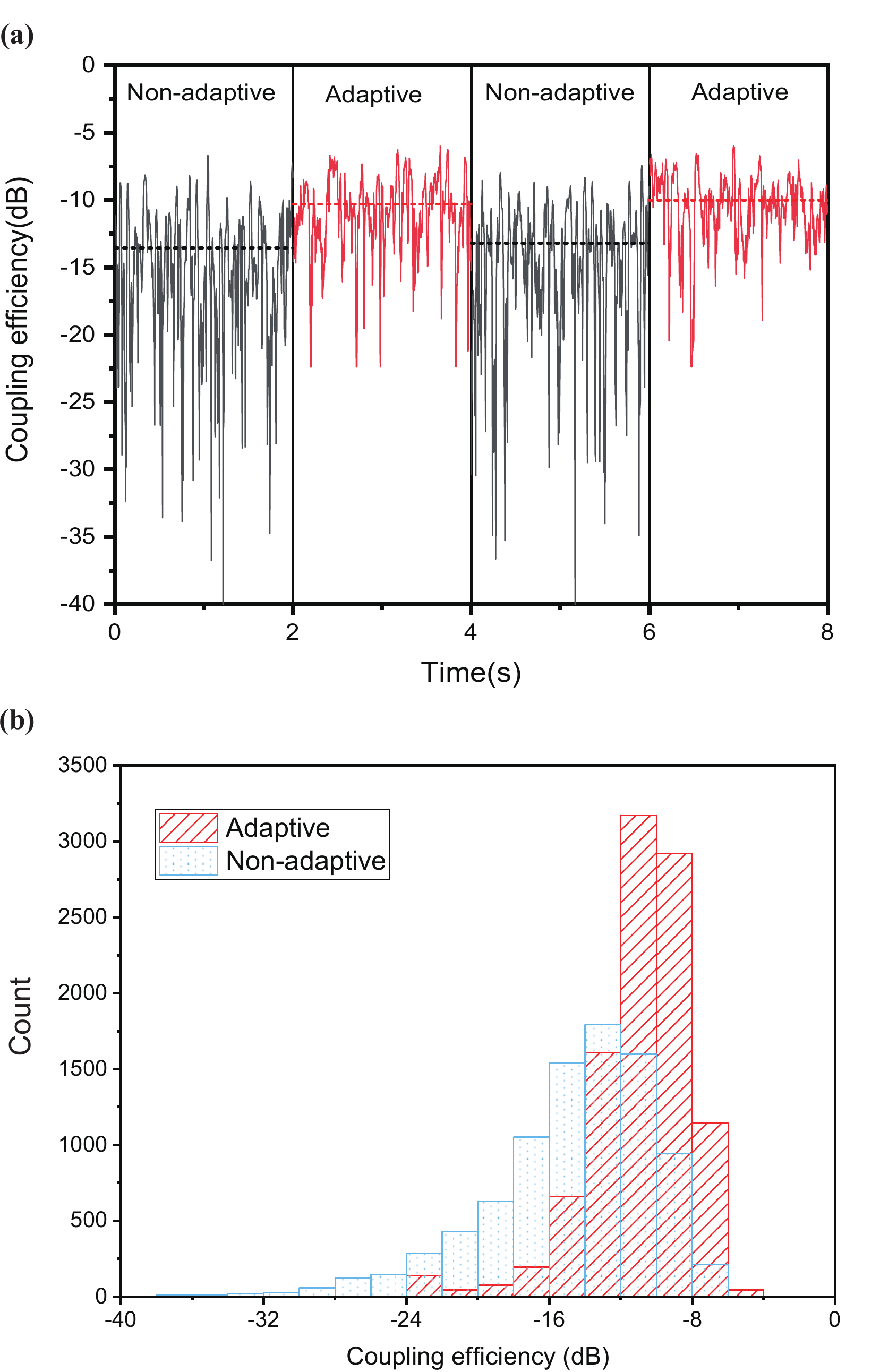}
\caption{\textbf{Performance of our AO system based on the M-SPGD algorithm under $ \bm{D/r_{0}=5.4} $ ( $ \bm{r_{0}=7.4} $ cm @810 nm).} \textbf{(a)}, Single-mode fiber coupling efficiency changes over time. The mean closed-loop single-mode coupling efficiency is approximately 3.1 dB higher than that of the open-loop. \textbf{(b)}, Probability distribution of the coupling efficiency. The closed-loop single-mode coupling efficiency relative standard deviation (RSD) is approximately 51.1\% while the open-loop single-mode coupling efficiency RSD is approximately 88.7\%.}
\label{Fig. 4}
\end{figure}
\par
As shown in Fig. 5(a), for a normalized atmospheric coherence length of $ D/r_{0}=9.5 $ ($ r_{0} $=4.2 cm @810 nm), which corresponds to strong atmospheric turbulence, the improvement of single-mode coupling efficiency is about 3.7 dB from open-loop coupling efficiency $ 3.0\% $ to closed-loop coupling efficiency $ 7.6\% $ by using the M-SPGD AO system and the single-mode coupling efficiency RSD significantly reduced from open-loop $ 97.3\% $ to closed-loop $ 53.7\% $. The utility of M-SPGD algorithm for improving single-mode coupling efficiency and suppressing under strong turbulence is also verified. From the histogram of coupling efficiency statistical distribution Fig. 5(b), similar to moderate intensity turbulence situation, the closed-loop coupling efficiency gathers to a smaller span. The open-loop and closed-loop refer to the status of AO system, and the fine APT system is always a closed-loop state.
\begin{figure}[htbp]
\centering
\includegraphics[width=9cm]{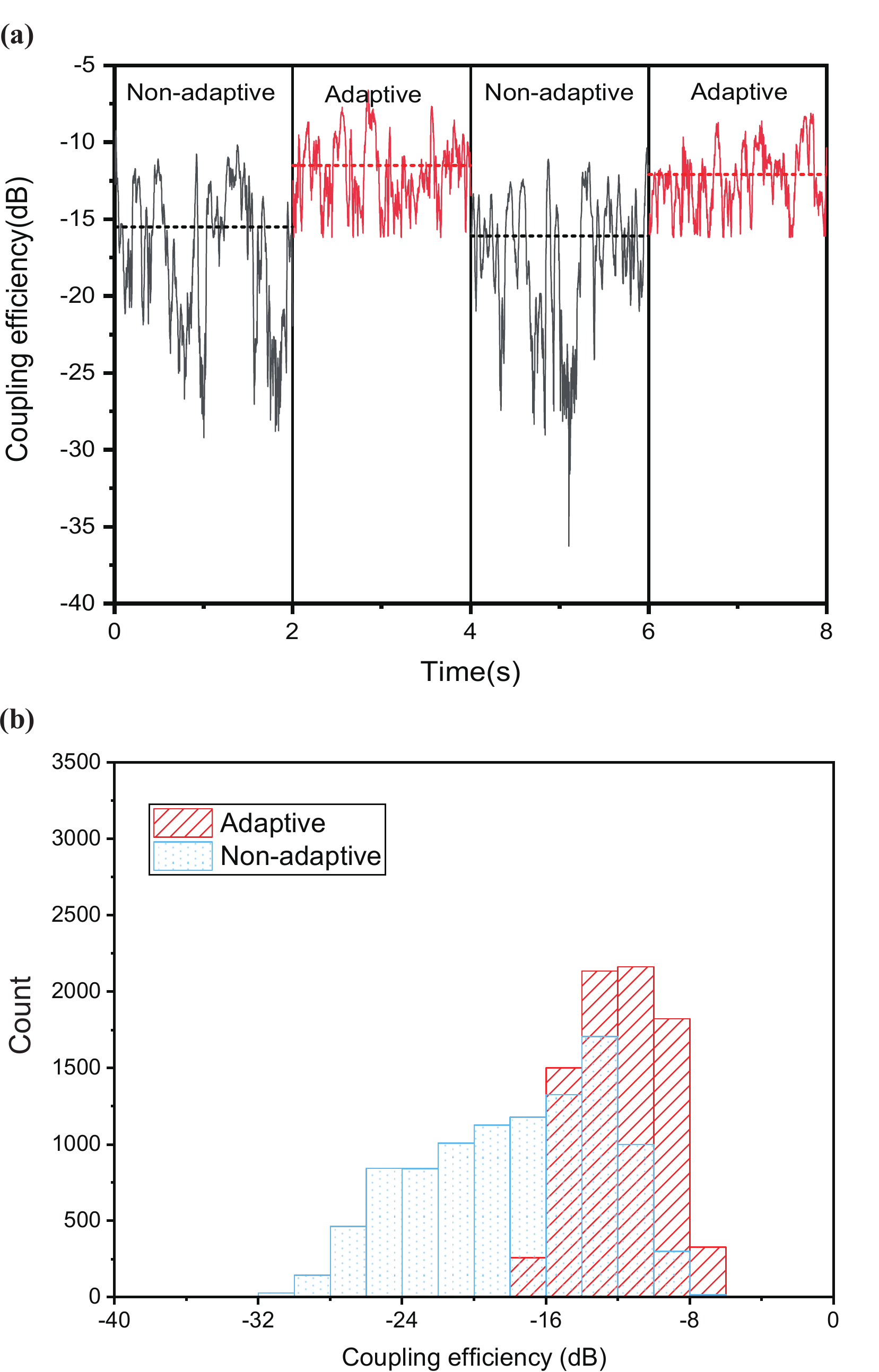}
\caption{\textbf{Performance of our AO system based on the M-SPGD algorithm under $ \bm{D/r_{0}=9.5} $ ($ \bm{r_{0}} $=4.2 cm @810 nm).} \textbf{(a)}, Single-mode fiber coupling efficiency changes over time. The mean closed-loop single-mode coupling efficiency is approximately
3.7 dB higher than that of the open-loop. \textbf{(b)}, Probability distribution of the coupling efficiency. The closed-loop single-mode coupling efficiency relative standard deviation (RSD) is approximately 53.7\% while the open-loop single-mode coupling efficiency RSD is approximately 97.3\%.}
\label{Fig. 5}
\end{figure}
\par
During normal operation of the M-SPGD AO system, the necessary emitting power of 1570 nm light is less than one microwatt. After passing through filter module, dark counts introduced by the 1570 nm light are measured to be less than 10 Hz, which is ignorable compared with detector noise and solar background noise. So, the M-SPGD AO system we developed can effectively improve the signal-to-noise ratio and is of great value in free-space quantum communications. 
\par
Due to low signal-to-noise ratio during the daytime, satellite-based quantum communication is limited to nighttime at the present stage. In the future, global-scale and all-day quantum communication based on medium and high orbit satellites can be expected and the main challenge is to obtain a sufficient signal-to-noise ratio. The technique we developed can improve the signal-to-noise ratio under different atmospheric turbulence intensities and can be applied to future satellite-based quantum communication. Specifically, selection of the performance metric of the M-SPGD algorithm requires further consideration because of the demand for optical power.
\section{Conclusion}
We developed an optimized AO system based on an M-SPGD algorithm to optimize SMF coupling and implemented field test of the performance of the M-SPGD AO system over an 8 km urban free-space link for the first time. Under different atmospheric turbulence intensities, the M-SPGD AO system can improve SMF coupling efficiency and suppress efficiency fluctuations effectively, particularly under strong atmospheric turbulence ($ D/r_{0}=9.5 $, $ r_{0}=4.2 $cm @810 nm), about 3.7 dB single-mode coupling efficiency improvement was obtained and the coupling efficiency RSD reduced from 97.3\% to 53.7\%. Our experimental results demonstrate great application potential of the M-SPGD AO technology in long-range quantum communications. 
\par
Note that the bandwidth of our M-SPGD AO system is 500 Hz and a deformable mirror with 40 units is used. The performance of the system can be further improved by increasing bandwidth and using a deformable mirror with more units, because higher bandwidth leads to higher wave-front aberration compensation speed and more units lead to higher wave-front aberration compensation precision. 

\section*{Funding}
This work was supported by the National Key R\&D Program of China (Grants No. 2017YFA0303900), the National Natural Science Foundation of China (Grants No.U1738201, U1738142, 11654005, 11904358, 61625503, 11822409, and 11674309), the Chinese Academy of Sciences (CAS), Shanghai Municipal Science and Technology Major Project (Grant No.2019SHZDZX01), and Anhui Initiative in Quantum Information Technologies. Y.Cao was supported by the Youth Innovation Promotion Association of CAS (under Grant No.2018492).

\section*{Acknowledgments}
We acknowledge Sheng-Kai Liao, Han-Ning Dai, Bo Li for their helpful discussions during the course of this article.

\section*{Disclosures}
The authors declare no conflicts of interest.

\bibliography{manuscript}






\end{document}